\documentclass[onecolumn,draftcls]{IEEEtran}
\usepackage{mathrsfs}
\usepackage{amsfonts}
\usepackage{graphicx}
\usepackage{subfigure}

\newtheorem{theorem}{Theorem}

\newtheorem{lemma}[theorem]{Lemma}

\newcommand{\suppress}[1]{}

\newcommand{\SDee}{{\cal D}}

\newcommand{\Dee}{{\bf D}}
\newcommand{\bX}{{\bf X}}
\newcommand{\Del}{{\Delta_n}}
\newcommand{\bx}{{\bf x}}
\newcommand{\bY}{{\bf Y}}
\newcommand{\by}{{\bf y}}
\newcommand{\eX}{{\cal X}}
\newcommand{\eY}{{\cal Y}}
\newcommand{\eW}{{\cal W}}
\newcommand{\bU}{{\bf U}}

\newcommand{\ca}{c}
\newcommand{\cb}{c}

\newcommand{\cd}{c}
\newcommand{\ce}{c}

\newcommand{\cg}{c}

\newcommand{\ci}{c}
\newcommand{\cj}{c}

\newcommand{\cl}{c}

\newcommand{\co}{c}

\newcommand{\Real}{\mathbb{R}}

\newcommand{\Field}{\mathbb{F}}

\newcommand{\qed}{\hfill{$\Box$}}

\newcommand{\Bx}{{\bf x}}
\newcommand{\By}{{\bf y}}

\newcommand{\defn}{\stackrel{\triangle}{=}}

\begin{document}
\title{``Real'' Slepian-Wolf Codes}

\author{Bikash Kumar Dey, Sidharth Jaggi, and Michael Langberg}


\markboth{Dey, Jaggi, and Langberg: ``Real'' Slepian-Wolf Codes}{}
\maketitle
\thispagestyle{empty}

\footnotetext{The work in this paper was presented in part in ISIT 2008, Toronto, Canada, July 2008 \cite{ShenviDJL:08}.\\
B.~K.~Dey is with the Department of Electrical Engineering, Indian Institute of Technology Bombay, Mumbai, India, 400 076, email: bikash@ee.iitb.ac.in . \\
S.~Jaggi is with the Department of Information Engineering, Chinese University of Hong Kong, Shatin, N.T., Hong Kong, email: jaggi@ie.cuhk.edu.hk  \\
M. Langberg is with the Computer Science Division, Open University of Israel, 108 Ravutski St., Raanana 43107, Israel, email: mikel@openu.ac.il}

\begin{abstract}
We provide a novel achievability proof of the Slepian-Wolf theorem for i.i.d.
sources over finite alphabets. We demonstrate that random codes that are
linear over the real field achieve the classical Slepian-Wolf rate-region. 
For finite alphabets we show that typicality decoding is equivalent to
solving an integer program. Minimum entropy decoding is also shown to
achieve exponentially small probability of error.  The techniques used may
be of independent interest for code design for a wide class of information
theory problems, and for the field of compressed sensing.
\end{abstract}

\section{Introduction}

A well-known result by Slepian and Wolf in~\cite{SleW:73} characterizes
the rate-region for near-lossless source coding of distributed sources.
The result demonstrates that if two (or more) sources possess correlated
data, even independent encoding of the sources' data can still achieve
essentially the same performance as when the sources encode jointly.
This result has important implications for information theoretic
problems as diverse as sensor networks~\cite{PraKR:02},
secrecy~\cite{CsiszarN:00},
and low-complexity video encoding~\cite{PurR:02}.
Unfortunately for the distributed source coding problem, codes that
are provably both rate-optimal and computationally efficient to implement are
hard to come by. Section~\ref{sec:back} gives a partial history of
results for the Slepian-Wolf (SW) problem.

In this work we provide novel codes that asymptotically achieve
the SW rate-region with vanishing probability of error.
Our encoding procedure comprises of random linear operations
over the real field $\Real$, and
are hence called {\em Real Slepian-Wolf Codes} or RSWCs.
In contrast most other codes in the literature operate over appropriate
finite fields $\Field_q$.
We demonstrate that RSWCs can be used in a way that enables
the receiver to decode the sources' information by solving a set of
integer programs (IPs). Besides being interesting in their own
right as a new class of codes achieving the SW rate-region,
the relation between RSWCs and IPs has some
intriguing implications.

In general IPs are computationally intractable
to solve. However, our code design gives us significant flexibility
in choosing the particular IPs corresponding to our codes. That is,
we show that ``almost all'' RSWCs result in IPs that
have ``good'' performance for the SW problem. But there are
well-studied classes of IPs that are known to be computationally
tractable to solve (for e.g., IPs corresponding to {\em Totally Unimodular
matrices}~\cite{Hof:79}). It is thus conceivable that suitably
chosen RSWCs may be decodable with low computational complexity.

Linear SW codes over finite fields were introduced in \cite{Csi:82} and
they were shown to achieve the SW rate-region. Decoding such codes is
equivalent to finding a
vertex of a hypercube satisfying some combinatorial properties.
Such problems are computationally intractable.
Our SW codes are linear over $\Real$. Though decoding our codes may
still be difficult, we can use tools from the matured field of convex
optimization for decoding our codes.

Also, our work has direct implications for the new field of
{\em Compressed Sensing (CS)}.
In the CS setup, $N$ sources each generate a single real number.
The resulting length-$N$ sequence is {\em $k$-sparse}, i.e., can be
written with at most $k \ll N$ non-zero coefficients in a prespecified basis.
A typical result \cite{CanRT:06} in this setup shows that if a receiver gets
${\cal O}(k\log(N))$ random linear combinations over $\Real$ of the sources'
sequence, it can, with high probability,
reconstruct the source sequence exactly in a computationally
efficient manner by solving a linear program. The CS setup is quite similar
to that of the RSWCs we design --
the source sequence contains a large amount of redundancy, and
a random $\Real$-linear mixture of the sequence suffices for exact
reconstruction via optimization techniques.
There are, however, two major differences. First, RSWCs operate at
information-theoretically optimal rates whereas CS codes are bounded
away from such performance. Second, CS codes are computationally tractable,
whereas we are currently not aware of efficient decoding techniques for
RSWCs. We think this tradeoff between computational efficiency and
rate-optimality is interesting and worthy of further investigation.

In Section \ref{sec:back}, we discuss some background and tools
to be used in the subsequent sections. In Section \ref{sec:rswc},
we present the construction of our RSWCs and the related main results.
These results are then proved in Sections \ref{sec:thm:main} and \ref{sec:thm:main2}.
In Section \ref{sec:no_time_share}, we present the direct construction of RSWCs for any point
on the Slepian-Wolf rate-region without time-sharing between the
corner points. The universal minimum-entropy decoding algorithm is
shown to work for our RSWCs in Section \ref{sec:universal}. Section \ref{sec:nsn} shows
that our RSWCs achieve the rate-region of more general {\em normal
source networks without helpers} introduced in \cite{CsiK:80}. Finally
Section \ref{sec:conclusion} concludes the paper.

\section{Background and Definitions}\label{sec:back}

Shannon's seminal source coding theorem~\cite{Shannon:48} demonstrates
that a sequence of discrete random
variables can essentially be compressed down to
the entropy of the underlying probability distribution generating the sequence.
Of the many extensions sparked by this paper, the Slepian-Wolf
theorem~\cite{SleW:73} is the one this paper builds on.

\subsection{Slepian Wolf Theorem for i.i.d. sources~\cite{SleW:73}}
\noindent {\bf Problem Statement:} Two sources named Xavier and Yvonne
generate two sequences of discrete random
variables, $\bX \defn X_1, X_2, \ldots, X_n$ over the finite alphabet $\eX$, and
$\bY \defn Y_1, Y_2, \ldots, Y_n$ over the finite alphabet $\eY$, respectively.
The sequence $(\bX,\bY)$ is assumed to
be i.i.d. with a joint distribution $p_{X,Y}(x,y)$ that is known
in advance to both Xavier and Yvonne. The
corresponding marginal distributions over $X$ and $Y$ are denoted by $p_X(x)$ and $p_Y(y)$
respectively.
Xavier and Yvonne wish to communicate $(\bX,\bY)$ to
a receiver Zorba.
To this end Xavier uses his
encoder
to transmit a message
that is a function only of $\bX$ and $p_{X,Y}(x,y)$ to Zorba.
Similarly, Yvonne uses her encoder
to transmit a message
that is a function only of $\bY$ and $p_{X,Y}(x,y)$ to Zorba.
Zorba uses his decoder to attempt to reconstruct $(\bX,\bY)$.
Xavier and Yvonne's encoders and Zorba's decoder comprise a SW code ${\cal C}$.
The SW code ${\cal C}$
is said to be {\em near-lossless}
if Zorba's reconstruction of
$(\bX,\bY)$ is correct with a probability of error over $p_{XY}(x,y)$
that is asymptotically negligible
in the {\em block-length} $n$. The {\em rate-pair} $(R_X, R_Y)$
is said to be {\em achievable} for the SW problem if for every $\epsilon > 0$
there exists a code ${\cal C}$ that
is near-lossless, and the average (over $p_{X,Y}(x,y)$)
number of bits that ${\cal C}$ requires
Xavier and Yvonne to transmit to Zorba are at most $n(R_X + \epsilon)$ and
$n(R_Y + \epsilon)$
respectively. The set of all rate-pairs that are achievable is called the
{\em rate-region}.
Slepian and Wolf's characterization of the rate-region is remarkably clean.
\begin{theorem}{\cite{SleW:73}}
The rate-region for the
Slepian-Wolf problem is given by the intersection of
\begin{eqnarray}
R_X & \geq & H(X|Y), \nonumber\\
R_Y & \geq & H(Y|X),\label{eq:sw_reg}\\
R_X+R_Y & \geq & H(X,Y).\nonumber
\end{eqnarray}
\end{theorem}
Here $H(X|Y)$ and $H(Y|X)$ denote the {\em conditional entropy} and
$H(X,Y)$ denotes the {\em joint entropy} of $(X,Y)$ (implicitly,
over the joint distribution $p_{X,Y}(x,y)$).

\subsection{Linear SW codes over finite fields}
The SW codes in~\cite{SleW:73} have
computational complexity that is exponential
for both encoding and decoding. An improvement was made in~\cite{Csi:82},
where it was shown that {\em random linear} encoders suffice.
We briefly restate that result here, restricting ourselves to
the case when $\eX = \eY = \{0,1\}$ for simplicity.

Let $\Dee_\bX$ and $\Dee_\bY$ be respectively $\lceil n(R_X + \epsilon)\rceil \times n$ and
$\lceil n(R_Y + \epsilon) \rceil \times n$ matrices over the finite field $\Field_2$,
with each
entry of both matrices chosen i.i.d. as either $0$ or $1$
with probability $1/2$. Here $\epsilon$ is an arbitrary positive constant.
Abusing notation, let $\bX$ and $\bY$ also denote length-$n$ column vectors
over $\Field_2$.
Xavier and Yvonne's encoders are then defined respectively via the matrix
multiplications
$\Dee_\bX\bX$ and $\Dee_\bY\bY$, and their messages to Zorba are respectively
the resulting column vectors.

We now define Zorba's decoder. For an arbitrary distribution
$p_{X,Y}(x,y)$ over finite alphabets, let the {\em strongly}
$\epsilon$-{\em jointly typical set}
$A_{\epsilon,p_{X,Y}}^n$~\cite{CoverT:91} (henceforth simply called
the {\it typical} set) be the set of all length-$n$ sequences
$({\bX},{\bY})$ such that the empirical distribution induced by
$({\bX},{\bY})$ differs component-wise from $p_{X,Y}(x,y)$ by at
most $\epsilon/(|\eX||\eY|)$. That is,
\begin{eqnarray}
&& A_{\epsilon,p_{X,Y}}^n \defn \left\{ (\bx, \by) :
\left|\frac{N_{(\bx,\by)}(a,b)}{n} - p_{X,Y}(a,b)\right|
< \frac{\epsilon}{|\eX||\eY|} \mbox{ for every } (a,b) \in
\eX \times \eY \right\} \nonumber
\end{eqnarray}
where $N_{(\bx,\by)}(a,b)$ denotes the number of component pairs $(x_i, y_i)$
in $(\bx, \by)$ which are equal to $(a,b)$.
For simplicity of notation we denote
$A_{\epsilon,p_{X,Y}}^n$ as $A_\epsilon$. Zorba checks to see if
there exists a unique length-$n$ sequence $(\hat{\bX},\hat{\bY})$
satisfying two conditions. First, that $\Dee_\bX\hat{\bX}$ and
$\Dee_\bY\hat{\bY}$ respectively match the messages transmitted by
Xavier and Yvonne. Second, whether $(\hat{\bX},\hat{\bY})$ lies
within $A_\epsilon$. If both conditions are satisfied for exactly
one sequence $(\hat{\bX},\hat{\bY})$, Zorba outputs
$(\hat{\bX},\hat{\bY})$, else he declares a decoding error.

Then~\cite{Csi:82} shows the following result.
\begin{theorem}{\cite{Csi:82}}
For each rate pair $(R_X, R_Y)$ in the region defined by (\ref{eq:sw_reg})
and sufficiently large $n$, with high
probability over choices of $\Dee_\bX$ and $\Dee_\bY$ the corresponding
SW code is near-lossless.
\end{theorem}

Many of the SW codes in the literature build on such encoders that are
linear over a finite field. Some such codes use iteratively
decodable channel codes
to attain performance that is empirically ``good'', but performance
guarantees have not been proven
(e.g.~\cite{GarZ:01}). Other codes use recent
theoretical advances in channel codes to produce near-lossless codes that
achieve any point in the SW rate-region, but cannot give guarantees on
computational complexity (e.g.~\cite{ColLME:04}).

\subsection{Linear codes over real fields}

As mentioned in the introduction, Compressed Sensing codes operate over
real (and complex) fields, and are structurally
similar to the codes proposed in this work. The primary difference between
the two sets of results is that our focus is
on achieving information-theoretically optimal performance
(at the cost of potentially high decoding complexity), whereas CS codes
have lower decoding complexity at the cost of non-optimal rates.
Some intriguing results on CS codes can be found in~\cite{Don:06,CanRT:06}.

Concurrently, codes over the real field $\Real$ also seem to have applications for the channel coding problem.
Using significantly different techniques, Tao et al.~\cite{CanT:05}
obtained
channel codes that can be decoded solving a linear program (LP). Also, {\em lattice codes}
have been shown to achieve capacity for the AWGN channel~\cite{UrbR:98}.

\section{RSWC Model}
\label{sec:rswc}

As is common in the SW literature~\cite{CoverT:91}, we focus
on just the point $(H(X),H(Y|X))$ in the SW rate-region. Time-sharing
between this and the symmetric point $(H(X|Y),H(Y))$ enables us to
achieve all points in the rate-region.
Thus Xavier encodes his data $\bX$ using a
classical lossless source code, and Zorba decodes it losslessly.
We henceforth discuss only Yvonne's RSWC encoder
for $\bY$ and Zorba's corresponding decoder.
In Section~\ref{sec:no_time_share} we show how to generalize our
proof techniques to get codes that achieve any point in the SW rate-region without time-sharing.
We consider only $\eX$ and $\eY$ that are ordered finite subsets of $\Real$.

\noindent {\bf RSWC Encoder:} We define an
$\Real^{m\times n}$ {\em encoding
matrix} $\Dee$. Here $m$ is a code-design parameter to be specified
later, and $\Dee$ is chosen as follows. Each component $D_{ij}$
of $\Dee$ is chosen randomly from a finite set $\SDee$. More
precisely, each element of $\Dee$ is chosen i.i.d. from $\SDee$
according to a distribution $p_D$. The set $\SDee$ can be any
arbitrary finite subset of $\Real$, and the distribution $p_D$
can be chosen arbitrarily on $\SDee$, as long as the probability of
at least two elements of $\SDee$ is non-zero.
For ease of proof, we assume that $p_D$ is zero-mean -- the more general case
requires only small changes in the proof details.
The particular values
of $\SDee$ and $p_D$ can be chosen according to the application.
We denote the $i$-th row of $\Dee$ by $\Dee_i$.

For a fixed block-length $n$, Yvonne's data is arranged as a column
vector $\bY
\defn (Y_1, Y_2, \cdots, Y_n)^T$.
To encode, $\bY$ is multiplied by $\Dee$ to get a length-$m$ real
vector ${\bf U} \defn \Dee \bY$. We denote the real interval
$(-n^{0.5+\epsilon}, n^{0.5+\epsilon})$ by $I_q$. Each component $U_i$ of
${\bf U}$ is uniformly quantized by dividing $I_q$ into steps of
size $\Delta_n=2n^{-\epsilon}$. Thus $\lceil (0.5+2\epsilon)\log n \rceil$ bits suffice for this
quantization. Note that the values outside the range $I_q$ are quantized
to the farthest quantization levels from origin.
Here and throughout the paper $\log(.)$ denotes the binary
logarithm, and
$\epsilon$ is a code-design parameter that can be
used to trade off between the probability of error and the rate of
the RSWC. It can be chosen as any arbitrarily small positive real
number. The quantized value of $U_i$ is denoted by $\hat{U}_i$
and the corresponding length-$m$ quantized vector is denoted
by $\hat{\bf U}$. We take $m = \left\lceil (n(H(Y|X)+3\epsilon))/(0.5\log n) \right\rceil$
since then Yvonne's encoder will encode at about $H(Y|X)$ bits per symbol.
Thus the total number of bits Yvonne transmits to Zorba equals $m\lceil (0.5+2\epsilon)\log n \rceil$,
which for all sufficiently large $n$ can be bounded from above by $nH(Y|X) + \rho\epsilon n$ for
a universal constant $\rho$.


\noindent {\bf RSWC Decoder:}
Zorba first decodes ${\bf X}= \bx$. Suppose he
received $\hat{\bf U}=\hat{\bf u}$ from Y. He finds a vector ${\bf
y}$ which is strongly $\epsilon$-jointly typical with $\bx$,
and for which
$\widehat{\Dee\by} = \hat{\bf u}$. If there is
no such $\by$ or there is more than
one such $\by$ he declares a decoding error.

The ensemble of RSWC encoder-decoder pairs described above is denoted by
${\cal C}(\epsilon,n,p_{X,Y}, p_D)$. The {\em probability of error of}
${\cal C}(\epsilon,n,p_{X,Y}, p_D)$ is defined as the probability over $p_{X,Y}$
and $p_{\Dee}$ that
Zorba makes or declares a decoding error. The {\em rate of}
${\cal C}(\epsilon,n,p_{X,Y}, p_D)$ is defined as the
number of bits that Yvonne transmits to Zorba.

We are now in a position to state and prove our main results.
The proofs of these results are presented in the next two sections.
Theorem \ref{thm:main} shows that our RSWCs achieve the corner
point $(H(X),H(Y|X))$ in the Slepian-Wolf rate-region with exponentially
small probability of error.
\begin{theorem}
For all sufficiently large $n$ there are universal positive constants $c,\rho$,
such that the probability of error under typicality decoding and rate
of ${\cal C}(\epsilon,n,p_{X,Y}, p_D)$
are at most $2^{-cn/\log n}$ and $H(Y|X) + \rho\epsilon $ respectively.
\label{thm:main}
\end{theorem}

We next show that Yvonne's decoding can be done by solving an IP.

\begin{theorem}
If Yvonne's source is binary, then the typicality decoding of a RSWC
for the point $(H(X), H(Y|X))$ is equivalent to solving an IP.
\label{thm:main1}
\end{theorem}

Further, we show that even for discrete memoryless sources over larger alphabet
$\eY$, the
encoder can be implemented as a series of RSWC encoders 
each of which is for a derived binary source. Then the typicality decoder can
be implemented as a series of decoders each of
which is equivalent to solving an IP.

\begin{theorem}
For any finite alphabet $\eY$, the real SW encoding can be done using
$|\eY|-1$ RSWC encoders so that the typicality decoder can be implemented by solving
$|\eY|-1$ IPs.
\label{thm:main2}
\end{theorem}

For any rate-pair in the Slepian-Wolf rate-region,
a direct construction of the individual RSWC encoders for Xavier and Yvonne
without time-sharing between the corner points is presented in
Section \ref{sec:no_time_share}. It is shown that RSWCs constructed this
way also achieve the Slepian-Wolf rate-region.

\begin{theorem}
Any point in the Slepian-Wolf rate-region can be achieved directly
by RSWCs without time-sharing.
\label{thm:no_time_sharing}
\end{theorem}

We also show that RSWCs can be decoded by {\em minimum
entropy decoding}.

\begin{theorem}
For all sufficiently large $n$ there are universal positive constants $c,\rho$,
such that the probability of error under minimum entropy decoding and rate
of ${\cal C}(\epsilon,n,p_{X,Y}, p_D)$
are at most $2^{-cn/\log n}$ and $H(Y|X) + \rho\epsilon $ respectively.
\label{thm:main3}
\end{theorem}

It is argued in Section \ref{sec:nsn} that the achievable rate-region of the
more general class of source networks known as {\em normal source networks
without helpers}~\cite{CsiK:80} is also achieved by our RSWCs.

\begin{theorem}
Random RSWCs achieve the rate region of any normal source network
without helpers.
\label{thm:main4}
\end{theorem}

The above results will be proved in the subsequent sections.
In the rest of the paper, for simplicity of exposition many different constants, independent of $n$,
will be denoted by the same symbol ``$c$''.

\section{Proof of Theorem~\ref{thm:main}}\label{sec:thm:main}
The probability of decoding error is given by
\begin{eqnarray}
P_e^n \leq P_1 + P_2 \label{eq:pe}
\end{eqnarray}
where
$P_1$ is the probability that $(\bX, \bY)$ are not
strongly jointly $\epsilon$-typical, and $P_2$ is the probability that
$(\bX, \bY) \in A_\epsilon$, but there
is another $\by^\prime \neq \by$ such that
$(\bX, \by^\prime) \in A_\epsilon$, and $\widehat{\Dee\bY} =
\widehat{\Dee\by^\prime}$.

\noindent
{\bf Bounding $P_1$:} For $P_1$, note that for any non-typical sequence $(\bx, \by)$,
its type $p_{(\bx, \by)}$ satisfies $|p_{X,Y} - p_{(\bx, \by)}|_1
\geq \epsilon/|\eX||\eY|$. So, using $D(p_{X,Y}||p_{(\bx, \by)})
\geq |p_{X,Y} - p_{(\bx, \by)}|_1^2/(2\ln 2 )$ 
\cite[Lemma 12.6.1]{CoverT:91} and Sanov's theorem
\cite[Theorem 12.4.1]{CoverT:91}, we have
\begin{eqnarray}
P_1 & \leq & (n+1)^{|\eX||\eY|} \exp\left(-n. \frac{\epsilon^2}{2 |\eX|^2|\eY|^2} \right) \nonumber \\
& \leq & 2^{-\cg n} \label{eq:p1}
\end{eqnarray}
for some positive constant $\cg$. The rest of this section focuses on bounding
$P_2$ in (\ref{eq:pe}).

\begin{figure}[h]
\centering
\includegraphics[width=2.5in]{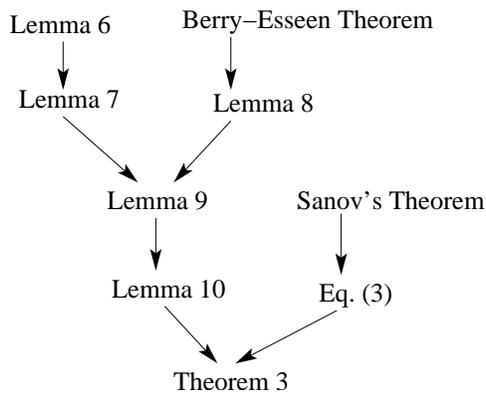}
\caption{Dependence structure of Lemmas}
\label{fig:dependence}
\end{figure}

\noindent
{\bf Bounding $P_2$:}
In the following, we present a sequence of
lemmas leading to Lemma \ref{lem:qerr}, which gives a bound on $P_2$.
A dependency ``graph'' of lemmas is shown in Fig. \ref{fig:dependence}
to ease understanding. We start by a general lemma proved in the Appendix.

\begin{lemma}
Let $W_1, W_2, \cdots, W_n$ be a sequence of i.i.d. zero-mean random variables
taking values from $\eW$, and $a \defn \max \{|w| | w \in \eW\}$. Then
for any positive constant $A$,
$$
Pr \left\{\left|\sum_{i=1}^n W_i \right| > A \right\} \leq 2 (n+1)^{|\eW |} \exp \left(-\frac{A^2}{2na^2 }\right)
$$
\label{lem:ld}
\end{lemma}


We now show some properties of our quantization of $U_i = \Dee_i \bY$.
\begin{lemma}
There exists a positive constant $c$ so that for any $\by \in \eY^n$,
$$
\Pr\{|\Dee_i \by |> n^{0.5+\epsilon }\} \leq 2^{-cn^{2\epsilon}}.
$$
\label{lem:ld2}
\end{lemma}
{\bf Proof:}
Let $y_{\max}$ be the element in $\eY$ with maximum absolute value.
For any $y\in \eY$, let $S_y$ be the set of indices $j$ such that $y_j = y$,
i.e., $S_y \defn \{j|y_j = y\}$. 
If $|\Dee_i \by | = \left|\sum_{y\in \eY}
\left(\sum_{j \in S_y} D_{ij}y_j\right)\right| > n^{0.5+\epsilon }$ then
for at least one $y$, $|\sum_{j \in S_y} D_{ij}y_j| >
(1/|\eY |) n^{0.5+\epsilon}$. 
So,
\begin{eqnarray}
\lefteqn{\Pr\{|\Dee_i \by |> n^{0.5+\epsilon }\}} \nonumber \\
& \leq &
\Pr\left\{\left|\sum_{j \in S_y} D_{ij}y_j\right| > \frac{1}{|\eY |} n^{0.5+\epsilon}
\mbox{ for at least one } y\right\} \nonumber \\
&\leq & \sum_{y\in {\cal Y}} \Pr\left\{\left|\sum_{j \in S_y} D_{ij}y_j\right| > \frac{1}{|\eY |} n^{0.5+\epsilon}\right\} \nonumber \\
& = & \sum_{y\in {\cal Y}} \Pr\left\{ \left|\sum_{j \in S_y} D_{ij}\right| > \frac{1}{|\eY |
|y|}n^{0.5+\epsilon}\right\} \nonumber \\
&\leq & \sum_{y\in \eY} \left\{ 2(|S_y|+1)^{|{\cal D}|}
\exp\left({-\frac{1}{2|S_y|\alpha^2 }\frac{1}{|\eY|^2|y|^2} n^{1+2\epsilon }}\right)
\right\} \label{eq:ld2.1} \\
&\leq & \sum_{y\in \eY}  2(n + 1)^{|{\cal D}|}
\exp\left({-\frac{n^{1+2\epsilon }}{2n\alpha^2 |\eY|^2|y_{\max}|^2} }\right) \label{eq:ld2.2} \\
&\leq & |\eY |  2(n + 1)^{|{\cal D}|}
\exp\left({-\frac{n^{2\epsilon}}{2\alpha^2 |\eY|^2|y_{\max}|^2}}\right) \nonumber \\
&\leq & 2^{-cn^{2\epsilon }} \nonumber
\end{eqnarray}
for some constant $c$, for large enough $n$, and
where $\alpha = \max \{|d| | d \in {\cal D}\}$.
Here (\ref{eq:ld2.1}) follows from Lemma \ref{lem:ld}, and
(\ref{eq:ld2.2}) follows from $|S_y|\leq n$ and $|y_{\max}| \geq |y|\,\,
\forall y \in \eY$.
\qed


The following lemma gives, for two different $\by, \by^\prime \in
\eY^n$, an upper bound on the probability that $\widehat{\Dee_i{\bf
y}} = \widehat{\Dee_i\by^\prime}$.


Let $p_\pm$ denote the minimum of $Pr\{D_{ij}>0\}$ and
$Pr\{D_{ij}<0\}$. Since $D_{ij}$ has zero mean and has at least two
symbols with non-zero probability, it follows that $p_\pm \neq 0$.
\begin{lemma}
If $\by \in \eY^n$ and $\by^\prime \in \eY^n$ differ in $t$
components then
$$
Pr\{|\Dee_i(\by - \by^\prime)|< \Del\} \leq
\min \left(1-p_\pm, \frac{\cl}{\sqrt{t}}\right)
$$
for some fixed constant $\cl \in \Real$.
\label{lem:be}
\end{lemma}
{\bf Proof:}
Let $b_y$ be the {\em smallest difference in} $\eY$, i.e.,
$b_y \defn \min_{y_1, y_2 \in \eY, y_1\neq y_2} |y_1-y_2|$.
We denote the $j$-th component $(y_j-y_j^\prime)$ of $\by -\by^\prime$
by $\alpha_j$. Then there are $t$
nonzero $\alpha_j$, and w.l.o.g., we assume that
$\alpha_1, \alpha_2, \ldots, \alpha_t \neq 0$. 
Note that $|\{y-y^\prime | y, y^\prime \in \eY, y\neq y^\prime\}| \leq
|\eY|^2$. So there are at least $\tau \defn t/|\eY|^2$ elements among $\alpha_1,
\alpha_2, \ldots, \alpha_t$ which are the same. Let us assume, w.l.o.g.,
that $\alpha_1 = \alpha_2 =\cdots = \alpha_{\tau}$. Let $\sigma^2$ be the
variance of $D_{ij}$. Then the random variables
$V_1=\alpha_1D_{i1}, V_2=\alpha_2 D_{i2}, \ldots, V_\tau =\alpha_\tau D_{i\tau}$
are i.i.d. with zero mean and variance $\sigma^{\prime 2} = |\alpha_1 |^2 \sigma^2$.
The central limit theorem states that the distribution of the normalized
sum $W_\tau=\sum_{j=1}^\tau V_j / (\sigma^\prime \sqrt{\tau})$ approaches
the normal ${\cal N}(0,1)$ distribution as $\tau$ increases. The
{\it Berry-Esseen theorem}~\cite{Feller:72}
gives a uniform upper bound on the deviation of the cumulative
distribution function (cdf) of $W_\tau$ from the cdf of ${\cal N}(0,1)$.
The Berry-Esseen bound is given by
\begin{eqnarray}
|Pr\{W_\tau < w\} - \Phi(w) | \leq \frac{\beta \gamma}{{\sigma^\prime}^3\sqrt{\tau}},
\label{eq:be}
\end{eqnarray}
for any $w\in {\mathbb{R}}$. Here $\gamma = E\{|V_1|^3\}$ is the
third moment of $V_1$, and $\beta$ is a universal constant whose
value has been improved over the decades.
We use the Berry-Esseen bound to prove the lemma as below.
\begin{eqnarray}
\lefteqn{Pr\{|\Dee_i(\by-\by^\prime)|<\Del\}} \nonumber \\
& & =
 Pr\{-\Del <\Dee_i(\by-\by^\prime) < \Del\}\nonumber \\
&  & = Pr\left\{-\frac{\Del}{|\alpha_1| \sigma \sqrt{\tau}} <\frac{\Dee_i(\by-\by^\prime)}{|\alpha_1| \sigma \sqrt{\tau}} < \frac{\Del}{|\alpha_1| \sigma \sqrt{\tau}}\right\}\nonumber \\
& & \leq  Pr\left\{-\frac{\Del}{\sigma b_y \sqrt{\tau}} <\frac{\Dee_i(\by-\by^\prime)}{|\alpha_1| \sigma \sqrt{\tau}} < \frac{\Del}{\sigma b_y \sqrt{\tau}}\right\}\nonumber \\
& & =  Pr\left\{-\frac{\sum_{j=\tau +1}^nD_{ij}(y_j-y_j^\prime)}{|\alpha_1| \sigma \sqrt{\tau}} -\frac{\Del}{\sigma b_y \sqrt{\tau}} 
 < \frac{\sum_{j=1}^\tau D_{ij}(y_j-y_j^\prime)}{|\alpha_1| \sigma \sqrt{\tau}} 
< - \frac{\sum_{j=\tau +1}^nD_{ij}(y_j-y_j^\prime)}{|\alpha_1| \sigma \sqrt{\tau}} + \frac{\Del}{\sigma b_y \sqrt{\tau}}\right\} \nonumber \\
& & =  Pr\left\{-\frac{\sum_{j=\tau +1}^nD_{ij}(y_j-y_j^\prime)}{|\alpha_1| \sigma \sqrt{\tau}} -\frac{\Del}{\sigma b_y \sqrt{\tau}} 
 <  W_\tau
< - \frac{\sum_{j=\tau +1}^nD_{ij}(y_j-y_j^\prime)}{|\alpha_1| \sigma \sqrt{\tau}} + \frac{\Del}{\sigma b_y \sqrt{\tau}}\right\}\label{BE1}\\
& & \leq  \frac{2\Del}{\sigma b_y \sqrt{\tau}\sqrt{2\pi}} + 2\times \frac{\beta \gamma}{{\sigma^\prime}^3\sqrt{\tau}}
\label{BE} \\
&  &  = \frac{\cl}{\sqrt{t}} \nonumber
\end{eqnarray}
Eq. (\ref{BE}) follows by using the Berry-Esseen bound (\ref{eq:be})
on the normalized sum $W_\tau$.
The first term $2 \times \frac{1}{\sqrt{2\pi }}
\times \frac{\Delta_n}{\sigma b_y \sqrt{\tau}}$ 
in (\ref{BE}) is an upper bound on the probability
of ${\cal N}(0,1)$ lying in the interval of length
$2\times \frac{\Delta_n}{\sigma b_y \sqrt{\tau}}$ in (\ref{BE1}). This bound
is obtained by multiplying the maximum value $1/\sqrt{2\pi}$ of the
probability density function of ${\cal N}(0,1)$ by the length of the interval.
The deviation of the cdf of $W_\tau$
from that of ${\cal N}(0,1)$ at each boundary point of the interval
is bounded by the Berry-Esseen bound. The second term in (\ref{BE})
is the sum of this bound at these two boundary points.

For $t>0$, there is at least one $j$ such that $y_j \neq
y_j^\prime$. Let us assume, w.l.o.g., that $y_1 \neq y_1^\prime$.
For large enough $n$, $\Del < b_y \times \min_{d \in \SDee, d\neq 0}
 |d|$. So,
\begin{eqnarray}
Pr\{|\Dee_i(\by-\by^\prime)|<\Del\} \leq 1-p_\pm . \nonumber
\end{eqnarray}
This can be easily checked by considering the
change in the value from $\sum_{j=2}^n D_{ij}(y_j-y_j^\prime)$
to $\Dee_i(\by-\by^\prime )$.
\qed

\begin{lemma}
Let $\by$ and $\by^\prime$ be any two vectors differing
in $t$ components. Then for some constant $c$ and a constant
$\tilde{p} < 1$, both independent of $\by$ and $\by^\prime$, we have
$$
\Pr\{\widehat{\Dee_i\by}= \widehat{\Dee_i\by^\prime}\}
\leq \min \left(\tilde{p}, \frac{\cl}{\sqrt{t}}\right) 
$$
for large enough $n$.
\label{lem:qerr}
\end{lemma}
{\bf Proof:}
\begin{eqnarray}
\lefteqn{\Pr\{\widehat{\Dee_i\by}= \widehat{\Dee_i\by^\prime}\}} \nonumber \\
&& \leq  \Pr\{\widehat{\Dee_i\by}= \widehat{\Dee_i\by^\prime} |
|\Dee_i\by | \leq n^{0.5+\epsilon}, |\Dee_i\by^\prime | \leq n^{0.5+\epsilon}\}
+ \Pr\{|\Dee_i\by | > n^{0.5+\epsilon}\} + \Pr\{|\Dee_i\by^\prime | > n^{0.5+\epsilon}\}\nonumber \\
&& \leq  \Pr\{|\Dee_i(\by - \by^\prime)|< \Del\} 
+ \Pr\{|\Dee_i\by | > n^{0.5+\epsilon}\} + \Pr\{|\Dee_i\by^\prime | > n^{0.5+\epsilon}\}\nonumber \\
&& \leq  \min \left(1-p_\pm, \frac{\cl}{\sqrt{t}}\right) + 2(2^{-cn^{2\epsilon }}) \label{eq:ld} 
\end{eqnarray}
for large enough $n$. The second term in (\ref{eq:ld}) is obtained by
applying Lemma \ref{lem:ld} on the last two terms in the previous line.
For any constant $c^\prime > c$, we have $\frac{\cl}{\sqrt{t}} + 2(2^{-cn^{2\epsilon }})
< \frac{c^\prime}{\sqrt{t}} $ for large enough $n$. Also, for any
$\tilde{p} > 1-p_\pm$, $1-p_\pm + 2(2^{-cn^{2\epsilon }}) < \tilde{p}$ for
large enough $n$. So the result follows.
\qed

We are now ready to present an upper bound on $P_2$.
\begin{lemma}
For large enough $n$,
\begin{eqnarray}
P_2 \leq 2^{-\ce n/\log n}, && \label{eq:p2}
\end{eqnarray}
where $\ce$ is a constant.
\label{lem:p3}
\end{lemma}
{\bf Proof:}
\begin{eqnarray}
P_2
& = & \hspace*{-0mm} \sum_{(\bx, \by) \,\in
\,A_\epsilon } p_{X,Y}(\bx, \by)
Pr\left\{\exists \by^\prime \neq \by \mbox{ s. t. }
\widehat{\Dee\by^\prime} = \widehat{\Dee\by},
 (\bx, \by^\prime) \,\in A_\epsilon
\right\}
\nonumber \\
& \leq & \hspace*{-0mm}\sum_{(\bx, \by) \,\in
\,A_\epsilon }
\hspace*{-0mm}p_{X,Y}(\bx, \by)
\hspace*{-0mm}\mathop{\sum_{
\by^\prime \neq \by}}_{(\bx, \by^\prime) \,\in A_\epsilon}
\hspace*{-0mm} Pr\left\{
\widehat{\Dee\by^\prime} = \widehat{\Dee\by} \right\}
\label{eq:p2.3} \\
& = & \hspace*{-0mm}\sum_{(\bx, \by) \,\in A_\epsilon}
p_{X,Y}(\bx, \by) \sum_{t>0}
\hspace*{-0mm}\mathop{\sum_{(\bx, \by^\prime) \,\in
 A_\epsilon}}_{d_H(\by, \by^\prime) = t}
\hspace*{-0mm}\left(Pr\{\widehat{\Dee_1\by^\prime} = \widehat{\Dee_1\by} \}
\right)^m \label{eq:p2.4} \\
& \leq & \sum_{(\bx, \by) \,\in A_\epsilon} p_{X,Y}(\bx, \by) \sum_{t>0} \hspace*{-0mm}\mathop{\sum_{(\bx, \by^\prime) \,\in A_\epsilon}}_{d_H(\by, \by^\prime) = t}
\hspace*{-0mm}\left(\min \left(\tilde{p}, \frac{\cl}{\sqrt{t}}\right)\right)^m \label{eq:p2.5} \\
& = & \sum_{(\bx, \by) \,\in A_\epsilon} p_{X,Y}(\bx, \by) \sum_{t>0}
N_{\bx,\by}(t) \left(\min \left(\tilde{p}, \frac{\cl}{\sqrt{t}}\right)\right)^m \label{eq:p2.1}
\end{eqnarray}
where $N_{\bx, \by}(t)$ is the number of $\by^\prime$ which are jointly
typical with $\bx$ and which are at Hamming distance $t$ from $\by$, i.e.,
$N_{\bx, \by}(t) \defn |\{\by^\prime \in \eY^n|(\bx,
\by^\prime) \in A_\epsilon, d_H(\by, \by^\prime) = t \}|$.
Eq. (\ref{eq:p2.3}) follows by union bound, Eq. (\ref{eq:p2.4}) follows
because the rows of $\Dee$ are i.i.d., and Eq. (\ref{eq:p2.5}) follows
from Lemma \ref{lem:qerr}.
For $t>0$, let $N(t)$ denote the maximum of $N_{\bx, \by}(t)$ over all possible
typical $(\bx, \by)$ pairs, i.e., $N(t) \defn \max_{(\bx, \by)
\in A_\epsilon} N_{\bx,\by}(t)$. Further, let $t_n$ denote the
value of $t$ for which the expression inside the second summation in
(\ref{eq:p2.1}) takes the maximum value for some typical $(\bx, \by)$,
i.e., $t_n \defn \arg \max_{t>0} \left(N(t)
\left(\min \left(\tilde{p}, \cl/\sqrt{t}\right)\right)^m\right)$.
The subscript in $t_n$ is to emphasize that it is a function of $n$.
Then by substituting in (\ref{eq:p2.1}),
\begin{eqnarray}
P_2 & \leq & nN(t_n)  \left(\min \left(\tilde{p}, \frac{\cl}{\sqrt{t_n}}\right)\right)^m .\nonumber
\end{eqnarray}

We emphasize here that every appearance of ``$c$''
may denote a different constant in the following.

For any $\delta \leq \epsilon/2(H(Y|X)+3\epsilon )$,
we consider two regimes: (1) $t_n > n^{1-\delta}$ and (2)
$t_n \leq n^{1-\delta}$.
In the first regime, we use
the bounds $N(t_n) \leq 2^{n(H(Y|X)+2\epsilon)}$
\cite[Theorem 14.2.2]{CoverT:91},
$Pr\{\widehat{\Dee_i \by} \neq \widehat{\Dee_i \by^\prime}\} \leq \cl/\sqrt{t}$, and
$m = \left\lceil (n(H(Y|X)+3\epsilon))/(0.5\log n) \right\rceil$ to get,
for large enough $n$,
\begin{eqnarray}
\log (P_2) & \leq & \log n + \log N(t_n) 
-\frac{n(H(Y|X)+3\epsilon)}{0.5\log n}((0.5-0.5 \delta)\log n
-\log \cb ) \nonumber \\
& = & n(H(Y|X) + 2\epsilon) - n(H(Y|X) + 3\epsilon )(1-\delta) 
+ n\frac{(H(Y|X)+3\epsilon )}{0.5\log n} \cb + \log n \label{eq:p2.reg1} \\
& = & -n\left(\epsilon - \delta(H(Y|X) + 3 \epsilon)\right) 
+ n\left(\frac{(H(Y|X)+3\epsilon )}{0.5\log n} \cb + \frac{\log n}{n}\right). \nonumber
\end{eqnarray}
Now, using $\delta \leq \epsilon/2(H(Y|X)+3\epsilon )$ and
$\left(c(H(Y|X)+3\epsilon )/0.5\log n + (\log n)/n\right)
< \epsilon /4$ for sufficiently large $n$, we get
\begin{eqnarray}
\log (P_2) & \leq & -\frac{n\epsilon}{2} + \frac{n\epsilon}{4} \nonumber \\
& = & -\frac{n\epsilon}{4}. \label{eq:p2.reg1a}
\end{eqnarray}
In the regime $t_n \leq n^{1-\delta}$,
we use the bounds
$N(t_n) < (|\eY|-1)^{t_n} {n\choose t_n} < (|\eY|n)^{t_n}$, and
$Pr\{\widehat{\Dee_i \by} \neq \widehat{\Dee_i\by^\prime}\} \leq
\tilde{p}$ to get
\begin{eqnarray}
\log (P_2) & \leq & \log n + t_n\log n + t_n \log |\eY| - \frac{n(H(Y|X)+3\epsilon)}{\log n}\log \left(\frac{1}{\tilde{p}}\right)  \nonumber \\
& \leq &  \log n + n^{1-\delta}\log n + n^{1-\delta} \log |\eY| - \frac{\cd n}{\log n} , \label{eq:p2.reg2}
\end{eqnarray}
where $\cd = (H(Y|X)+3\epsilon )\log (1/\tilde{p})$.
For large enough $n$, $(\log n)^2 < \cd n^{(\delta /2)}/3
\Rightarrow \log n < \cd n^{(\delta /2)}/(3\log (n))$. Also,
for large enough $n$, $n^{-\delta} \log |{\cal Y}| < \cd/(3\log (n))$
for some constant $\cd$.
So, for some constant $c^\prime$,
\begin{eqnarray}
\log (P_2) & \leq & \log n - \frac{c^\prime n}{3 \log n}  \nonumber \\
 & \leq & - \frac{\ce n}{\log n}  \label{eq:p2.reg2a}
\end{eqnarray}
for large enough $n$ and for some constant $\ce$.

Since $\ce n/\log n < n\epsilon /4$ for large enough $n$,
the result follows by combining (\ref{eq:p2.reg1a}) and (\ref{eq:p2.reg2a}). \hfill $\Box$

From (\ref{eq:pe}), (\ref{eq:p1}), and (\ref{eq:p2}), we have, for large enough $n$,
$$
P_e^n \leq P_1 + P_2 \leq 2 P_2 \leq 2^{-c n/\log n},
$$
for a constant $c$, thus completing the proof of Theorem \ref{thm:main}.

\section{Proof of Theorem~\ref{thm:main1} and Theorem~\ref{thm:main2}}\label{sec:thm:main2}
We first show that for $\eY=\{0,1\}$ the typicality decoding of our scheme can be done via the solution of an IP.
Recall that for a vector $\by$, we defined, for any $y\in \eY$,
$S_y = \{i|y_i = y\}$.
Similarly with abuse of notation, for any vector $\bx = (x_1, \ldots, x_n)$
decoded by Zorba, and $x \in \eX$, let us define $S_x = \{i|x_i = x\}$.
The constraint $(\Bx,\By) \in A_\epsilon$
can be written as the linear constraints
$$
p(1,x) - \frac{\epsilon}{|\eX||\eY|} \leq \frac{1}{n}\sum_{i \in S_x}y_i \leq p(1,x) + \frac{\epsilon}{|\eX||\eY|}, \ \ \ \forall x \in \eX
$$
Moreover, the constraints $\widehat{\Dee \by}=\hat{{\bf u}}=(\hat{u}_1,\dots,\hat{u}_n)$ can be written as
$$
\hat{u}_i-\Del/2 \leq \Dee_i\by \leq \hat{u}_i+\Del/2, \ \ \ \forall i=1,\dots m.
$$
Finally we add the `integrality' constraints, namely, that $\By \in \eY^n$.

For arbitrary finite
alphabets $\eY$, Yvonne and Zorba perform $|\eY|-1$ encoding and decoding
stages, each of which involves IP decoding of a binary vector. A sketch follows.

Let $y^{(1)},\ldots,y^{(|\eY|)}$ denote the distinct values of $\eY$.
In the first stage, instead of encoding $\by$ directly, Yvonne uses ${\cal C}(\epsilon,n,p^1_{X,Y}, p_D)$
to encode the vector $f^1(\by)$. Here the vector $f^1(\by)$ equals $1$ in the locations that $\by$ equals $y^{(1)}$ and equals $0$ otherwise, and $p^1_{X,Y}$ is the corresponding induced distribution
$p_{X,f^1(Y)}$ defined on $\eX \times \{0,1\}$.
Since $f^1(\by)$ is a binary vector,
Zorba can use the IP decoding described above, and therefore can retrieve
the locations where $\by$ equals $y^{(1)}$.
Inductively, in the $i$th stage,
Yvonne uses ${\cal C}(\epsilon,n(i),p^i_{X,Y}, p_D)$ to encode the vector $f^i(\by)$.
Here $n(i)$ equals the number of locations whose values are still undetermined before the $i^{th}$ stage, i.e., $n(i)$ equals $|\{j| \by_j \geq y^{(i)}\}|$. The length-$n(i)$ vector $f^i(\by)$ is obtained by
first throwing away the locations in $f^{i-1}(\by)$ that equalled $1$, and then marking the remaining
locations $1$ if and only if the corresponding locations in $\by$ equal
$y^{(i)}$.
At each stage, Zorba can use the IP decoding described above, and therefore can retrieve
the locations where $\by$ equals $y^{(i)}$.
Let $f^i(Y)$ denote the corresponding binary random variable s. t.
$(X, f^i(Y))$ has the joint distribution given by $p^i_{X,Y}(x,1)
= Pr\{X=x,Y=y^{(i)}|Y \neq y^{(1)}, y^{(2)}, \ldots, y^{(i-1)}\}$,
and $p^i_{X,Y}(x,0)
= Pr\{X=x,Y\neq y^{(i)}|Y \neq y^{(1)}, y^{(2)}, \ldots, y^{(i-1)}\}$.
Then by a direct extension of the grouping axiom \cite[Page 8]{Ash:65}, we have 
\begin{eqnarray}
H(Y|X) &=& H(f^1(Y)|X) + (1-p_Y(y^{(1)})) H(f^1(Y)|X) + (1-p_Y(y^{(1)})-p_Y(y^{(2)})) H(f^2(Y)|X) + \ldots \nonumber \\
&& \hspace*{10mm} + (p_Y(y^{(|\eY|-1)}) + p_Y(y^{(|\eY|)})) H(f^{|\eY|-1}(Y)|X).
\label{eq:grouping}
\end{eqnarray}
Clearly, for a single stage encoding/decoding, the average codelength
for Yvonne is bounded by $nH(Y|X) + c\epsilon n$. For a multi-stage
encoding/decoding as described above,
for a typical $\by$, the block length at the $i$-th stage is bounded
by $n(i) \leq n (1- Pr\{Y\in \{y^{(1)}, y^{(2)}, \ldots, y^{(i-1)}\}\}
+\epsilon)$ and so the codelength is bounded as
\begin{eqnarray}
L_i & \leq & n (1- Pr\{Y\in \{y^{(1)}, y^{(2)}, \ldots, y^{(i-1)}\}\}
+\epsilon) H(f^i(Y)|X) + c_i \epsilon n \nonumber
\end{eqnarray}
for some constants $c_i$. The average codelength is thus bounded
using (\ref{eq:grouping}) by
\begin{eqnarray}
L \leq \sum_{i=1}^{|\eY|-1} L_i \leq nH(Y|X) + c\epsilon n
\label{eq:Lbound}
\end{eqnarray}
for some constant $c$. If $\by$ is not typical, then in the worst
case, the codelength $n(i)=n$ for each $i$. Then the overall codelength
is bounded by $L \leq c^\prime n$ for some constant $c^\prime$. Since
the probability of the non-typical set is exponentially small, the
overall average codelength is still bounded by (\ref{eq:Lbound}) for
some constant $c$.
Hence the overall rate of this multistage RSWC differs from $H(Y|X)$ by at most $c\epsilon $,
where $c$ is some constant dependent only on $p_{X,Y}$.

The overall probability of error can be bounded as
\begin{eqnarray}
P_e^n & \leq & P_1 + \sum_{i=1}^{|\eY|} P_{2,i},
\end{eqnarray}
where $P_1$ is the probability that the vector $\by$ is not strongly
typical, and $P_{2,i}$ is the conditional probability of error at the
$i$-th stage of decoding given that the vector $\by$ is strongly
$\epsilon$-typical and the decoding till the $(i-1)$-th stage is correct.
If $\by$ is strongly $\epsilon$-typical, then the codelength at the $i$-th stage is $n(i) \geq n \times \sum_{j=i}^{|\eY|}
(P_Y(y^{(i)}) - \epsilon/|\eY|) \geq n (P_Y(y^{(|\eY|)}) - \epsilon/|\eY|)$.
So,
\begin{eqnarray}
P_{2,i} & \leq & \exp \left(-\frac{c^\prime n(i)}{\log (n(i))}\right) \nonumber \\
& \leq & \exp \left(-\frac{c^\prime n(i)}{\log n}\right) \nonumber \\
& \leq & \exp \left(-\frac{c^\prime (P_Y(|\eY|) - \epsilon/|\eY|) n}{\log n}\right). \nonumber 
\end{eqnarray}
Since $P_1$ is also exponentially small, the overall probability of error
for the multistage encoding/decoding is bounded as
\begin{eqnarray}
P_e^n & \leq & \exp \left(-\frac{cn}{\log n}\right). \nonumber 
\end{eqnarray}
\qed

\section{Real SW coding without timesharing}\label{sec:no_time_share}
Any rate-pair in the SW rate-region can also be directly achieved by
RSWCs without timesharing between the schemes achieving the rate-pairs
$(H(X|Y), H(Y))$ and $(H(X), H(Y|X))$. Let $(R_1, R_2)$ be a
rate-pair in the SW rate-region. Let
$m_1 = \left\lceil(n(R_1+3\epsilon))/(0.5\log n)\right\rceil$ and
$m_2 = \left\lceil(n(R_2+3\epsilon))/(0.5\log n)\right\rceil$. Similar to the
encoding scheme of Yvonne described in Section III,
Xavier chooses
an $m_1 \times n$ encoder matrix $\Dee_1$ over ${\cal D}$ according to
a distribution $P_D$. Similarly Yvonne
chooses a random $m_2 \times n$ encoder matrix $\Dee_2$ over ${\cal D}$
according to the distribution $P_D$
\footnote{Our arguments go through even if the elements of $\Dee_1$ and
$\Dee_2$ are chosen from different sets ${\cal D}_1$ and ${\cal D}_2$
according to some distributions. We restrict to ${\cal D}_1 = {\cal D}_2$
and the same distribution for the elements of $\Dee_1$ and $\Dee_2$
for simplicity.}.
Xavier encodes the length-$n$ vector $\bX$ by quantizing each component of
$\bU_1 \defn \Dee_1 \bX$ uniformly in the range $I_q$
with step-size $\Del = 2n^{-\epsilon}$ to obtain the vector $\widehat{\bU}_1$.
Similarly, Yvonne encodes the length-$n$ vector $\bY$ by quantizing each
component of $\bU_2 \defn \Dee_2 \bY$ uniformly in the range $I_q$
with step size $\Del = 2n^{-\epsilon}$ to obtain the vector $\widehat{\bU}_2$.

Zorba finds a unique jointly strongly $\epsilon$-typical pair $(\bx, \by)$
so that $\widehat{\Dee_1 \bx} = \widehat{\bU}_1$ and $\widehat{\Dee_2 \by} = \widehat{\bU}_2$.
If there is no such pair, or if there are more than one such pair, then
the decoder declares an error. The probability of error can be bounded as
\begin{equation}
P_e^n \leq P_1 + P_{21} + P_{22} + P_{23},
\label{eq:pen}
\end{equation}
where $P_1$, as before, is the probability that $(\bX, \bY)$ is not jointly
strongly $\epsilon$-typical,
$P_{21}$ is the probability that there is a $\bx^\prime \neq \bX$ which
is also jointly strongly $\epsilon$-typical with $\bY$ and $\widehat{\Dee_1\bx^\prime} = \widehat{\bU}_1$,
$P_{22}$ is the probability that there is a $\by^\prime \neq \bY$ which
is also jointly strongly $\epsilon$-typical with $\bX$ and $\widehat{\Dee_2\by^\prime} = \widehat{\bU}_2$,
and $P_{23}$ is the probability that there is another jointly typical pair
$(\bx^\prime, \by^\prime)$ so that $\bx^\prime \neq \bX, \by^\prime \neq \bY,
\widehat{\Dee_1\bx^\prime} = \widehat{\bU}_1$ and
$\widehat{\Dee_2\by^\prime} = \widehat{\bU}_2$.
We now investigate all the terms in (\ref{eq:pen}).

Let $\Dee_{1,i}$ and $\Dee_{2,i}$ denote the $i$-th rows of the matrices
$\Dee_1$ and $\Dee_2$ respectively. Similarly as Lemma \ref{lem:qerr},
we have
\begin{eqnarray}
Pr\{\widehat{\Dee_{1,i}\bx } = \widehat{\Dee_{1,i}\bx^\prime}\},  Pr\{\widehat{\Dee_{2,i}\by} = \widehat{\Dee_{2,i}\by^\prime}\} 
& \leq & \min \left(\tilde{p}, \frac{c_{1}}{\sqrt{t}}\right), \nonumber
\end{eqnarray}
when each pair $\bx, \bx^\prime \in \eX^n$ and $\by, \by^\prime \in \eY^n$
differ in $t$ positions.

We define the following functions.
\begin{eqnarray}
&& m(R) \defn \left\lceil\frac{n(R+3\epsilon )}{0.5 \log n}\right\rceil, \nonumber \\
&& \phi_1 (h, R, \delta) \defn \log \left( n2^{n(h+2\epsilon)}\left(\frac{\ca}{\sqrt{n^{1-\delta}}}\right)^{m(R)}\right), \mbox{ and}\nonumber \\
&&\phi_2 (L,R,\delta) \defn \log
\left(n(Ln)^{n^{1-\delta}}(\tilde{p} )^{m(R)} \right). \nonumber
\end{eqnarray}
Note that in this notation, $P_2$ in Lemma \ref{lem:p3} is given by
\begin{eqnarray}
\log (P_2) \leq \phi_1 (H(Y|X), H(Y|X), \delta)
\end{eqnarray}
for $t_n > n^{1-\delta}$ (See (\ref{eq:p2.reg1})).
As shown in (\ref{eq:p2.reg1a}), this is at most $-n\epsilon /4$
for $\delta \leq \epsilon/2(H(Y|X)+3\epsilon )$
for large enough $n$.
It can be checked similarly that for $\delta \leq
\epsilon /2(R+3\epsilon)$, $R\geq h$, and large enough $n$,
$\phi_1 (h, R, \delta) \leq -n((R-h)+\epsilon /4)$. Likewise, for
$t_n \leq n^{1-\delta}$, it is shown (See (\ref{eq:p2.reg2})) that
\begin{eqnarray}
\log (P_2) \leq \phi_2 (|\eY |, H(Y|X), \delta),
\end{eqnarray}
which is at most $-\ce n/\log n$ (See (\ref{eq:p2.reg2a})). More generally, it can
be similarly proved that for any constants $L>0$ and $\delta >0$,
\begin{eqnarray}
&& \phi_2 (L, R, \delta) \leq - \frac{\ci (R, \epsilon )n}{\log n}
\nonumber
\end{eqnarray}
for some constant $\ci (R, \epsilon ) > 0$ and for large enough $n$.

By definition,
\begin{eqnarray}
P_{22}
& = & \hspace*{-0mm} \sum_{(\bx, \by) \,\in A_\epsilon} p_{X,Y}(\bx, {\bf
y}) Pr\left\{\exists \by^\prime \neq \by \mbox{ s. t. } 
\widehat{\Dee\by^\prime} = \widehat{\Dee\by},
 (\bx, \by^\prime) \in A_\epsilon
\right\}.
\nonumber
\end{eqnarray}
By similar arguments to those in the proof of Lemma \ref{lem:p3}, we have
$\log (P_{22}) \leq \phi_2(|\eY |, R_2, \delta)$ for $t_n \leq n^{1-\delta}$,
and $\log (P_{22}) \leq \phi_1(H(Y|X), R_2, \delta)$ for $t_n > n^{1-\delta}$.
Since $R_2\geq H(Y|X)$, it follows that for large enough $n$,
\begin{eqnarray}
&&\log (P_{22}) \leq - \frac{\ci (R_2, \epsilon )n}{\log n}. \label{eq:p32}
\end{eqnarray}
Similarly, for large enough $n$,
\begin{eqnarray}
&&\log (P_{21}) \leq - \frac{\ci (R_1, \epsilon )n}{\log n}. \label{eq:p31}
\end{eqnarray}
As in the proof of Lemma
\ref{lem:p3}, $P_{23}$ can be simplified to (\ref{eq:p33a}) below, 
\begin{eqnarray}
P_{23}
& = & \hspace*{-2mm} \sum_{(\bx, \by) \,\in A_\epsilon } p_{X,Y}(\bx, \by)
Pr\left\{\exists (\bx^\prime, \by^\prime) \mbox{ s. t. }
\bx^\prime \neq \bx, \by^\prime \neq \by, 
\widehat{\Dee_1\bx^\prime} = \widehat{\Dee_1\bx},
\widehat{\Dee_2\by^\prime} = \widehat{\Dee_2\by},
 (\bx^\prime, \by^\prime) \,\in A_\epsilon \right\} \nonumber \\
& \leq & \hspace*{-2mm}\sum_{ (\bx, \by) \,\in A_\epsilon }
\hspace*{-2mm}p_{X,Y}(\bx, \by)
\hspace*{-0mm}\mathop{\sum_{\bx^\prime \neq \bx, \by^\prime \neq \by}}_{
(\bx^\prime, \by^\prime) \,\in A_\epsilon}\hspace*{-0mm} Pr\left\{
\widehat{\Dee_1\bx^\prime} = \widehat{\Dee_1\bx},\,
\widehat{\Dee_2\by^\prime} = \widehat{\Dee_2\by} \right\}
\nonumber \\
& = & \hspace*{-2mm}\sum_{(\bx, \by) \,\in A_\epsilon}
p_{X,Y}(\bx, \by) \sum_{t_1>0,\, t_2>0}
\hspace*{-0mm}\mathop{\sum_{(\bx^\prime, \by^\prime) \,\in A_\epsilon}}_{
d_H(\bx, \bx^\prime) = t_1,\, d_H(\by, \by^\prime) = t_2} Pr\left\{
\widehat{\Dee_1\bx^\prime} = \widehat{\Dee_1\bx},\,
\widehat{\Dee_2\by^\prime} = \widehat{\Dee_2\by} \right\}
\nonumber \\
& = & \hspace*{-0mm}\sum_{(\bx, \by) \,\in A_\epsilon}
p_{X,Y}(\bx, \by) \sum_{t_1>0,\, t_2>0}
\hspace*{-0mm}\mathop{\sum_{(\bx^\prime, \by^\prime) \,\in A_\epsilon}}_{
d_H(\bx, \bx^\prime) = t_1,\, d_H(\by, \by^\prime) = t_2}
\hspace*{-0mm}\left(Pr\{\widehat{\Dee_{1,1}\bx^\prime} = \widehat{\Dee_{1,1}\bx} \} \right)^{m(R_1)}\left(Pr\{\widehat{\Dee_{2,1}\by^\prime} = \widehat{\Dee_{2,1}\by} \} \right)^{m(R_2)} \nonumber \\
& \leq & \sum_{(\bx, \by) \,\in A_\epsilon} p_{X,Y}(\bx, \by)
\sum_{t_1>0,\, t_2>0}
\hspace*{-0mm}\mathop{\sum_{(\bx^\prime, \by^\prime) \,\in
 A_\epsilon}}_{d_H(\bx, \bx^\prime) = t_1,
\,d_H(\by, \by^\prime) = t_2}
\hspace*{-0mm}\left(\min \left(\tilde{p}, \frac{\ca}{\sqrt{t_1}}\right)\right)^{m(R_1)} \left(\min \left(\tilde{p}, \frac{\ca}{\sqrt{t_2}}\right)\right)^{m(R_2)} \nonumber \\
& = & \sum_{(\bx, \by) \,\in A_\epsilon} p_{X,Y}(\bx, \by)
\sum_{t_1,\, t_2 >0} N_{\bx, \by}(t_1, t_2) Q_1^{m(R_1)} Q_2^{m(R_2)}.
\label{eq:p33a}
\end{eqnarray}
In (\ref{eq:p33a}), $Q_1 = \min \left(\tilde{p}, \ca/\sqrt{t_1}\right)$,
$Q_2 = \min \left(\tilde{p}, \ca/\sqrt{t_2}\right)$,
and $N_{\bx, \by}(t_1, t_2)$ is the number of jointly typical
$(\bx^\prime, \by^\prime)$ pairs such that
$\bx^\prime$ differs from $\bx$ at $t_1$ locations and
$\by^\prime$ differs from $\by$ at $t_2$ locations, that is, 
$N_{\bx, \by}(t_1, t_2) \defn |\{(\bx^\prime, \by^\prime) \in
\eX^n\times \eY^n | (\bx^\prime, \by^\prime) \in A_\epsilon,
d_H(\bx, \bx^\prime) = t_1, d_H(\by, \by^\prime) = t_2\}|$.
We define $N(t_1, t_2) \defn \max_{\bx, \by} N_{(\bx, \by) \,\in A_\epsilon}(t_1, t_2)$,
and $(t_{1,n}, t_{2,n})$ as the pair $(t_1, t_2)$ that maximizes
$\left(N(t_1, t_2)
Q_1^{m_1} Q_2^{m_2} \right)$, that is,
$(t_{1,n}, t_{2,n}) \defn \arg \max_{t_1,t_2 >0} \left(N(t_1, t_2)
Q_1^{m_1} Q_2^{m_2} \right)$. Then
$$
P_{23} \leq n^2N(t_{1,n}, t_{2,n})Q_1^{m_1}Q_2^{m_2}.
$$
For
$\delta < \epsilon /2(R_1 + R_2 + 3\epsilon )$, we consider four cases.

Case I: $t_{1,n} > n^{1-\delta}, t_{2,n} > n^{1-\delta}$. In this case,
using the bounds $N(t_{1,n}, t_{2,n}) \leq 2^{n(H(X,Y) + \epsilon )},
Q_1 \leq \ca/\sqrt{t_{1,n}}, Q_2 \leq \ca/\sqrt{t_{2,n}}$,
we have
\begin{eqnarray}
\log (P_{23}) & \leq & \phi_1 (H(X,Y), R_1+R_2, \delta) \nonumber \\
& \leq & -n(R_1+R_2-H(X,Y)+\epsilon/4) \nonumber \\
& \leq & -n\epsilon/4. \label{eq:p33_1}
\end{eqnarray}

Case II: $t_{1,n} \leq n^{1-\delta}, t_{2,n} \leq n^{1-\delta}$. In this case,
using the bounds $N(t_{1,n}, t_{2,n}) \leq \left(|\eX|n\right)^{t_{1,n}}
\left(|\eY|n\right)^{t_{2,n}},
Q_1 \leq \tilde{p}, Q_2 \leq \tilde{p}$, we have
\begin{eqnarray}
\log (P_{23}) & \leq & \phi_2 (|\eX |, R_1, \delta) +
                    \phi_2 (|\eY |, R_2, \delta)\nonumber \\
& \leq & - \frac{\ci (R_1, \epsilon)n}{\log n} - \frac{\ci (R_2, \epsilon)n}{\log n}\nonumber \\
& \leq & - \frac{\cj (R_1, R_2, \epsilon)n}{\log n}, \label{eq:p33_2}
\end{eqnarray}
where $\cj (R_1, R_2, \epsilon) = \ci (R_1, \epsilon)+\ci (R_2, \epsilon)$.

Case III: $t_{1,n} > n^{1-\delta}, t_{2,n} \leq n^{1-\delta}$. In this case,
using the bounds $N(t_{1,n}, t_{2,n}) \leq 2^{n(H(X|Y)+2\epsilon )}
\left(|\eY|n\right)^{t_{2,n}},
Q_1 \leq \ca/\sqrt{t_{1,n}}, Q_2 \leq \tilde{p}$, we have
\begin{eqnarray}
\log (P_{23}) & \leq & \phi_1 (H(X|Y), R_1, \delta) + \phi_2 (|\eY |, R_2, \delta)\nonumber \\
& \leq & -n(R_1-H(X|Y)+\epsilon/4) - \frac{\ci (R_2, \epsilon)n}{\log n}
\nonumber \\
& \leq & - \frac{\cj (R_1, R_2, \epsilon)n}{\log n}. \label{eq:p33_3}
\end{eqnarray}

Case IV: $t_{1,n} \leq n^{1-\delta}, t_{2,n} > n^{1-\delta}$: As in
Case III, we have
\begin{eqnarray}
\log (P_{23}) & \leq & - \frac{\cj (R_1, R_2, \epsilon)n}{\log n}. \label{eq:p33_4}
\end{eqnarray}

From (\ref{eq:p1}), (\ref{eq:pen}),
(\ref{eq:p32}), (\ref{eq:p31}), (\ref{eq:p33_1}), (\ref{eq:p33_2}), (\ref{eq:p33_3}), and (\ref{eq:p33_4}), we have,
$$
P_e^n \leq 2^{-\co n/\log n }
$$
for some constant $\co$.

\section{Universal decoding: Proof of Theorem \ref{thm:main3}}
\label{sec:universal}
An encoding or decoding operation is said to be universal in a class
of sources if the encoding/decoding operation can be chosen without
the knowledge of the exact source statistics in the class.
The encoding for RSWCs without time-sharing in Section \ref{sec:no_time_share}
results in universal encoding 
in the class of i.i.d. sources. The two encoders may choose to encode
at rates $R_1$ and $R_2$ and choose their encoding matrices randomly
without the knowledge of the distribution of either source.
The joint typicality decoding discussed earlier will be able to recover both
the sequences with exponentially small probability of error as long
as the rate pair $(R_1, R_2)$ lies in the Slepian-Wolf rate
region of the sources. However, though the encoders are universal,
the joint typicality decoding is not universal since it requires the
decoder to know the joint distribution of the sources.

In this section, we show that the well known universal minimum entropy
decoding (MED) \cite{CsiK:80} which does not need the joint distribution of the sources will also
be able to decode our code with exponentially small probability of error
provided $m_1 \geq \left\lceil n(R_1+4\epsilon )/(0.5\log n)\right\rceil$ and
$m_2 \geq \left\lceil n(R_2+4\epsilon )/(0.5\log n)\right\rceil$ for some
$(R_1, R_2)$ in the Slepian-Wolf rate-region of the sources.
Here, the decoder finds the pair $(\bx, \by)$ with minimum empirical
entropy which satisfies the conditions $\widehat{\Dee_1\bx} = \widehat{\bU}_1$
and $\widehat{\Dee_2\by} = \widehat{\bU}_2$. If there are
more than one such pair then the decoder declares a decoding error.

Before investigating the probability of error under minimum entropy decoding,
let us define a weakly $\epsilon$-typical vector $(\bx, \by)$ as one satisfying
\begin{eqnarray}
&& \left| \log_2 (p_{X,Y}^n (\bx,\by ) ) + nH(X,Y)\right| \leq n\epsilon, \nonumber \\
&&  \left| \log_2 (p_{X}^n (\bx ) ) + nH(X)\right| \leq n\epsilon, \mbox{ and} \nonumber \\
&&  \left| \log_2 (p_{Y}^n (\by ) ) + nH(Y)\right| \leq n\epsilon . \nonumber 
\end{eqnarray}
The set of weakly $\epsilon$-typical vectors will be denoted by $A_{\epsilon, weak}$.
A weakly $\epsilon$-typical vector $\bx$ (similarly $\by$) is defined as one satisfying
\begin{eqnarray}
&&  \left| \log_2 (p_{X}^n (\bx ) ) + nH(X)\right| \leq n\epsilon. \nonumber 
\end{eqnarray}
The properties of the weakly typical set may be found in \cite{CoverT:91}.

Let us denote the joint entropy of the {\em type} of a pair of vectors
$(\bx, \by)$ as $H(\bx, \by)$, the corresponding conditional
entropies as $H(\bx | \by)$ and $H(\by | \bx)$, and the individual
entropies of the vectors as $H(\bx)$ and $H(\by)$. The probability
of error of a minimum entropy decoder is bounded as
\begin{equation}
P_e^n(MED) \leq P_1^\prime + P_{21}^\prime + P_{22}^\prime + P_{23}^\prime
\label{eq:mddpe}
\end{equation}
where $P_1^\prime$ is the probability that $(\bX, \bY)$ is not jointly weakly
$\epsilon$-typical,
$P_{21}$ is the probability that there is a $\bx^\prime \neq \bX$ such
that $H(\bx^\prime, \bY) \leq H(\bX, \bY)$ and $\widehat{\Dee_1\bx^\prime}
= \widehat{\bU}_1$, $P_{22}$ is the probability that there is a
$\by^\prime \neq \bY$ such that $H(\bX, \by^\prime) \leq H(\bX, \bY)$
and $\widehat{\Dee_2\by^\prime} = \widehat{\bU}_2$,
and $P_{23}$ is the probability that there is another pair
$(\bx^\prime, \by^\prime)$ so that $\bx^\prime \neq \bX, \by^\prime \neq \bY,
\widehat{\Dee_1\bx^\prime} = \widehat{\bU}_1$,
$\widehat{\Dee_2\by^\prime} = \widehat{\bU}_2$ and
$H(\bx^\prime , \by^\prime) \leq H(\bX , \bY)$.
We will briefly discuss all the terms in (\ref{eq:mddpe}).

By definition, $P_1^\prime = Pr\{A_{\epsilon, weak}^c\}$.
Since the weakly $\epsilon$-typical set is a superset of the
strongly $\epsilon^\prime (\epsilon, p_{X,Y})$-typical set for some 
$\epsilon^\prime (\epsilon, p_{X,Y})$~\cite{Yeung:08}, $P_1^\prime$ can be bounded similar
to (\ref{eq:p1}) as
\begin{eqnarray}
P_1^\prime & \leq & 2^{-\cg n} \label{eq:p1prime}
\end{eqnarray}
where the constant $c$ depends on $p_{X,Y}$.

Following similar steps as the proof of Lemma \ref{lem:p3}, we have
\begin{eqnarray}
P_{22} & = &
\sum_{(\bx, \by) \,\in A_\epsilon} p_{X,Y}(\bx, \by) \sum_{t>0}
N_{\bx,\by}^\prime (t) \left(\min \left(\tilde{p}, \frac{\cl}{\sqrt{t}}\right)\right)^{m_2} \nonumber
\end{eqnarray}
where $N_{\bx, \by}^\prime (t) \defn |\{\by^\prime \in \eY^n|
H(\by^\prime |\bx) \leq H(\by |\bx), d_H(\by, \by^\prime) = t \}|$.
Now, let us define $N^\prime(t) \defn \max_{(\bx, \by)\in A_\epsilon}
N_{\bx,\by}^\prime (t)$ for $t>0$, and $t_n \defn \arg \max_{t>0} \left(N^\prime (t)
\left(\min \left(\tilde{p}, \cl/\sqrt{t}\right)\right)^{m_2}\right)$.
Then clearly,
\begin{eqnarray}
P_{22} & \leq & nN^\prime (t_n)  \left(\min \left(\tilde{p}, \frac{\cl}{\sqrt{t_n}}\right)\right)^{m_2} .\nonumber
\end{eqnarray}
Note that for a given weakly typical $\bx$, the condition
$(\bx, \by) \in A_{\epsilon, weak}$ implies
$H(\by|\bx) \leq H(Y|X) + 2\epsilon$.
So, $N_{\bx, \by}^\prime (t) \subseteq |\{\by^\prime \in \eY^n|
H(\by^\prime |\bx) \leq H(Y|X)+2\epsilon, d_H(\by, \by^\prime) = t \}|$.
So, we can use both the bounds
$N^\prime (t_n) \leq 2^{n(H(Y|X)+3\epsilon)}$ and
$N^\prime (t_n) \leq (|\eY|n)^{t_n}$ for large enough $n$.
Then it can be shown in the same way as in the proof of Lemma \ref{lem:p3}
that $P_{22} \leq \exp (-cn/\log n)$
for $m_2 \geq \left\lceil n(R_2+4\epsilon )/(0.5\log n)\right\rceil$.
Similarly it can be shown that $P_{21}, P_{23} \leq \exp \,(-cn/\log n)$
for large enough $n$ if $m_1 \geq \left\lceil
n(R_1+4\epsilon )/(0.5\log n)\right\rceil$ and
$m_2 \geq \left\lceil n(R_2+4\epsilon )/(0.5\log n)\right\rceil$ for
a rate pair $(R_1, R_2)$ in the Slepian-Wolf rate-region.
Since $P_1^\prime$ goes to zero exponentially as in (\ref{eq:p1prime}),
it follows that
\begin{eqnarray}
P_e^n (MED) & \leq & \exp \,(-cn/\log n) \nonumber
\end{eqnarray}
for large enough $n$ for some constant $c$.

\section{Generalization to other source networks: Proof of Theorem \ref{thm:main4}}
\label{sec:nsn}
\begin{center}
\begin{figure}[h]
\centering\includegraphics[width=2.5in]{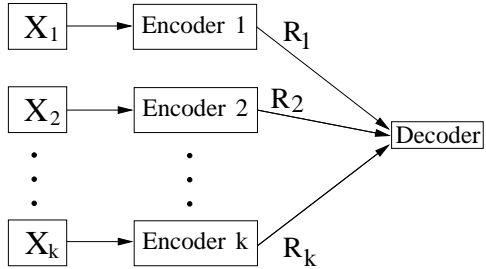}
\caption{A simple multi-source network}
\label{fig:multisource}
\end{figure}
\end{center}

The most simple generalization of the Slepian-Wolf source network
is to multiple sources as shown in Fig. \ref{fig:multisource}.
The same proof technique can be used to show that the decoder can
recover all the sources with exponentially small probability of error
if the encoders do random real encoding at rates satisfying
\begin{eqnarray}
&& \sum_{i\in {\cal L}} R_i \geq H(X_{\cal L}| X_{{\cal L}^c}) \nonumber
\end{eqnarray}
for each ${\cal L} \subseteq \{1,2,\cdots, k\}$. Here ${\cal L}^c$ denotes
the complement of ${\cal L}$. Using the same proof technique
as outlined in Sec. \ref{sec:universal}, one can show that the decoder
can also do minimum entropy decoding to attain vanishing probability
of error.
Csiszar and Korner \cite{CsiK:80} extended the result of Slepian and Wolf to
more general source networks called {\em normal source networks (NSN) without
helpers}. In the following, we briefly discuss their source network and
argue that our coding technique can achieve the achievable rate-region
of NSN without helpers.

\begin{figure}[h]
\centering
\includegraphics[width=2.5in]{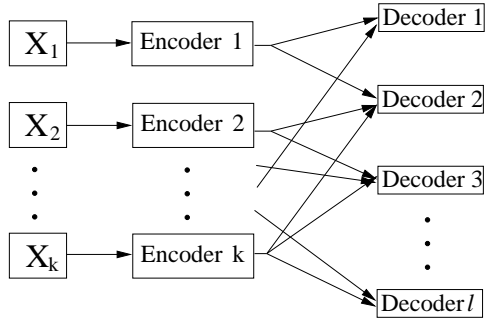}
\caption{Normal Source Network}
\label{fig:nsn}
\end{figure}

Let ${\cal A}$, ${\cal B}$ and ${\cal C}$ denote the
set of sources, encoders and decoders respectively in the network.
For any $c\in {\cal C}$, let ${\cal S}_c$ denote the set of source
nodes from which information is received at the decoder node $c$.
Let ${\cal D}_c$ denote the set of sources which are to be reproduced
at $c$.

An NSN, as defined in \cite{CsiK:80} and an example of which is shown
in Fig. \ref{fig:nsn}, is a source
network where

(i) there are no direct
edges from the sources to the decoders,

(ii) $|{\cal A}| = |{\cal B}|$ and
the edges from ${\cal A}$ to ${\cal B}$ define a one-to-one correspondence
between the sources and encoders,

(iii) all the sets ${\cal S}_c, c \in {\cal C}$ are different, and

(iv) for each pair of output vertices $c^\prime$ and $c^{\prime\prime}$,
the inclusion ${\cal S}_{c^\prime} \subseteq {\cal S}_{c^{\prime\prime}}$ implies
${\cal D}_{c^\prime} \subseteq {\cal D}_{c^{\prime\prime}}$.

For a source $a\in {\cal A}$, let $X_a$ denote the i.i.d. data generated
by the source. Similarly, for a subset ${\cal L} \subseteq {\cal A}$,
let $X_{\cal L}$ denote the vector $(X_a)_{a\in {\cal L}}$.
A source $a$ in an NSN is called a helper if for some $c\in {\cal C}$,
$a \in {\cal S}_c \setminus {\cal D}_c$. Clearly, a source network
without helpers satisfy ${\cal S}_c = {\cal D}_c$ for all $c \in {\cal C}$.
For any encoder $b\in {\cal B}$, let $R_b$ denote its encoding rate.
For a source network without helpers, Csiszar and Korner characterized
the rate-region. 

\begin{theorem}{\cite{CsiK:80}}
\label{th:nsn}
The achievable rate-region of an NSN without helpers equals the set
of those vectors $\bar{\bf R} = \{R_b\}_{b\in {\cal B}}$ which satisfy the
inequalities
\begin{eqnarray}
\sum_{b \in {\cal L}} R_b \geq H\left(X_{\cal L} | X_{{\cal S}_c\setminus{\cal L}}\right)
\end{eqnarray}
for every output $c\in {\cal C}$ and set ${\cal L} \subset {\cal S}_c$.
\end{theorem}

The achievability proof of this rate-region reduces to the achievability
proof of the corresponding rate-region for each of the networks
obtained by taking all the sources and one decoder. In other words,
if the encoders encode at rates satisfying the conditions
in Theorem \ref{th:nsn}, the probability of error for each
decoder is negligible. So the proof reduces to the proof
for the multiple source network as shown in Fig. \ref{fig:multisource}.
It thus follows that the rate-region of any NSN without helpers is
achievable by random real encoding at each encoder. Moreover,
the rate-region is also achievable with minimum entropy decoders.

\section{Conclusion}
\label{sec:conclusion}
The Real Slepian-Wolf Codes analyzed here provide a novel achievability proof of
the Slepian-Wolf theorem. Perhaps just as importantly, they demonstrate
the intriguing possibility of design of information-theoretic codes
via convex optimization techniques. For instance, since
decoding RSWCs is equivalent to solving an optimization problem, it is natural
to consider similar ``real'' codes for
problems where some function of the code simultaneously
needs to be optimized.
We are
currently investigating the performance of RSWCs under more structured choices of encoding matrices,
with the hope of obtaining codes for which IP decoding is equivalent to LP decoding, and is therefore
computationally tractable.

\appendix[Proof of Lemma \ref{lem:ld}]
First consider $Pr \left\{\sum_{i=1}^n W_i  > A \right\}$.
We define
$E \defn \{(w_1, w_2, \cdots, w_n)| \sum_{i=1}^{n} w_i > A\}$.
Let $p_w$ denote the
probability mass distribution of $W_i$. Then
\begin{eqnarray}
Pr \left\{\sum_{i=1}^n W_i  > A \right\}  & = & Pr\left\{E\right\} \nonumber \\
& = & Pr\left\{p_n|\mu_{p_n} > \frac{A}{n}\right\}. \nonumber
\end{eqnarray}
Here $p_n$ denotes the type of $(w_1, w_2, \cdots, w_n)$ and
$\mu_{p_n}$ denotes the mean of $p_n$. By Sanov's
Theorem~\cite[Theorem 12.4.1]{CoverT:91}, we have
\begin{eqnarray}
Pr \left\{\sum_{i=1}^n W_i  > A \right\} & = & p_{w}^n(E)
\leq (n+1)^{|\eW|} 2^{-nD(p_n^* || p_{w})},
\nonumber
\end{eqnarray}
where $p_n^* = \arg\min_{p_n: \mu_{p_n} > A/n}
D(p_n||p_{w})$. Since $p_w$ has zero mean, the ``nearest'' distribution
to $p_w$ that has mean greater than $A/n$ in absolute value would differ
from $p_w$ in the largest absolute component by at least $A/(an)$.
So, $\mu_{p_n^*}> A/n$ implies $|p_n^* - p_w|_1 > A/(an)$.
We then have
$D(p_n^* || p_{w}) \geq (1/2\ln 2) |p_n^* - p_w|_1^2
> A^2/(2 (na)^2 \ln 2 )$ by \cite[Lemma 12.6.1]{CoverT:91}. So,
\begin{eqnarray}
Pr \left\{\sum_{i=1}^n W_i  > A \right\} & \leq & (n+1)^{|\eW|} \exp\left(- \frac{A^2}{2 na^2 }\right).
\nonumber
\end{eqnarray}
Similarly one can show that $Pr \left\{\sum_{i=1}^n W_i  < -A \right\}  \leq  (n+1)^{|\eW|} \exp\left(- A^2/(2 na^2) \right)$. So the result follows.
\qed

\section*{Acknowledgments}
The authors gratefully acknowledge support from
the CUHK direct grant, the CU-MS-JL grant, and
a grant from the Bharti Centre for Communication.
We would like to thank S. Shenvi for his interest and involvement
in several stages of this work. We would also like to thank D. Manjunath
for fruitful discussions.

\bibliographystyle{ieeetr}
\bibliography{sid,diss}


\end{document}